\documentclass[%
reprint,
superscriptaddress,
nofootinbib,
amsmath,amssymb,
aps,
pra,
]{revtex4-2}

\usepackage{xcolor}

\usepackage{enumerate}
\usepackage{comment}
\usepackage{subfigure}
\usepackage{mathrsfs}
\usepackage{graphicx}
\usepackage{dcolumn}
\usepackage{bm}
\usepackage[colorlinks=true, allcolors=blue]{hyperref}

\usepackage{hyperref}
\usepackage{fancyhdr}
\usepackage[english=british]{csquotes}
\usepackage{amsmath}
\usepackage{amssymb}
\usepackage{physics}

\newcommand{\fullvec}[1]{\boldsymbol{\vec{#1}}}
\newcommand{\unitvec}[1]{\boldsymbol{\hat{#1}}}

\begin{document}

\title{Spin-current correlations in photoionization of chiral molecules}

\author{Philip Caesar M. Flores}
\affiliation{Max-Born-Institut, Max-Born-Str. 2A, 12489 Berlin, Germany}
\email{flores@mbi-berlin.de}
\author{Stefanos Carlstr\"om}
\affiliation{Max-Born-Institut, Max-Born-Str. 2A, 12489 Berlin, Germany}
\author{Serguei Patchkovskii}
\affiliation{Max-Born-Institut, Max-Born-Str. 2A, 12489 Berlin, Germany}
\author{Misha Ivanov}
\affiliation{Max-Born-Institut, Max-Born-Str. 2A, 12489 Berlin, Germany}
\affiliation{Insitute of Physics, Humboldt University zu Berlin, Berlin 12489, Germany}
\affiliation{Technion - Israel Institute of Technology, Haifa, Israel}
\author{Andres F. Ordonez }
\affiliation{Max-Born-Institut, Max-Born-Str. 2A, 12489 Berlin, Germany}
\affiliation{Department of Physics, Imperial College London, SW7 2BW London, United Kingdom}
\affiliation{Department of Physics, Freie Universit\"at Berlin, 14195 Berlin, Germany}
\author{Olga Smirnova}
\affiliation{Max-Born-Institut, Max-Born-Str. 2A, 12489 Berlin, Germany}
\affiliation{Technion - Israel Institute of Technology, Haifa, Israel}
\affiliation{Technische Universit\"at Berlin, 10623 Berlin, Germany}
\email{smirnova@mbi-berlin.de}
\date{\today}

\begin{abstract} 
Chirality-induced spin selectivity (CISS) refers to phenomena where molecular chirality governs spin polarization. While symmetry simply requires chiral molecules to support spin-vector correlations, we show that CISS is fundamentally a conditioned measurement of these correlations. We illustrate this principle for spin-resolved one-photon ionization of a randomly oriented ensemble of chiral molecules. We introduce and quantify the phenomenon of enantio-sensitive locking of the photoelectron current to its spin, thereby providing a complete description of spin-conditioned photoelectron currents in one-photon ionization.
\end{abstract}
\maketitle

Chirality-induced spin selectivity (CISS) originally refers to spin-polarized charge transport through chiral molecules \cite{ray1999asymmetric}, but more broadly denotes the remarkable ability of chiral structures to generate spin polarization in electron emission and transport processes \cite{CISS}. Since its discovery, CISS has triggered extensive debate regarding its microscopic origin and unexpectedly large magnitude \cite{naaman2015spintronics,bloom2024chiral,foo2025mind}, motivating detailed simulations \cite{gutierrez2012spin,zollner2020insight,naskar2023chiral}, minimal models \cite{eremko2013spin,gersten2013induced,fransson2020vibrational,ghazaryan2020analytic}, and new experimental probes \cite{mondal2016spin,eckvahl2023direct,albro2025measurement}. Experimentally, spin selectivity has been observed both in photoionization \cite{Goler2011,mollers2022chirality,naaman2012chiral} and in voltage-driven transport through chiral molecular films \cite{naaman2012chiral,naaman2020chiral,liu2023spin,Zhang2023}. In the latter case, a chiral molecular film is typically probed by magnetic conductive probe atomic force microscopy (mCP-AFM), where reversing the tip magnetization isolates the current conditioned on spin. 

These observations suggest that CISS-related phenomena are fundamentally linked to conditioned measurements, namely correlations $\langle \mathbf{v} \otimes \mathbf{s} \rangle$ between a vectorial molecular property $\mathbf{v}$ and the electron spin $\mathbf{s}$. Such pseudoscalar correlations are enabled by chiral structures and may manifest with different strengths depending on the measurement protocol. Starting from this perspective, we consider the simplest realization of chiral spin filtering, photoionization of an ensemble of randomly oriented chiral molecules, and identify the origin of the corresponding pseudoscalar correlations. In particular, we consider the case in which all external biases, such as molecular alignment or light polarization, are removed: a scenario that has not yet been examined elsewhere.

\begin{figure}[t!]
	\centering
	\includegraphics[width=0.22\textwidth]{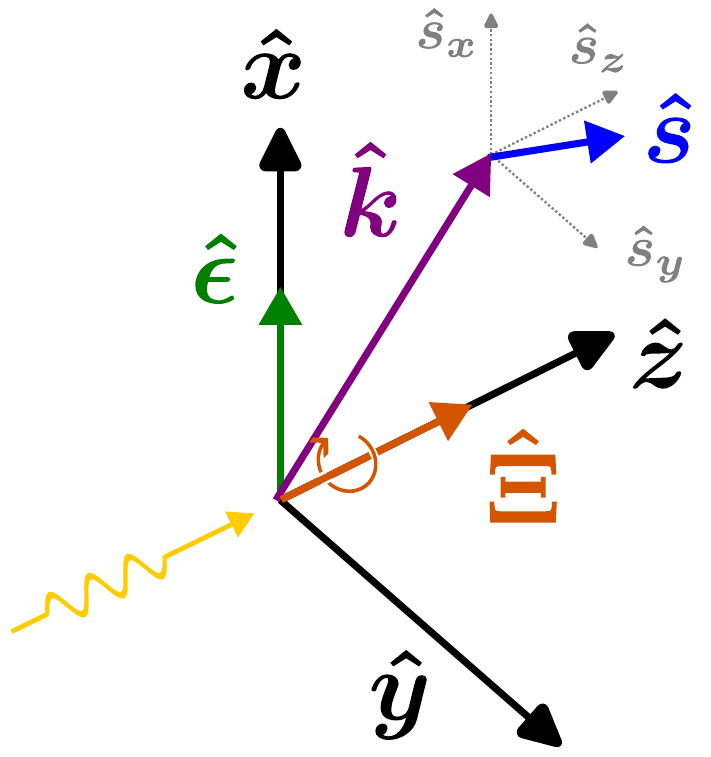}
	\caption{Specification of coordinates in the laboratory frame. Upon ionization, the photoelectron is ejected in the direction of $\unitvec{k}$ with its spin measured parallel to $\unitvec{s}$. The light field propagates along $\unitvec{z}$. The unit vector $\unitvec{\epsilon}$ denotes the polarization direction for linearly polarized light, while $\unitvec{\Xi}$ denotes the direction of photon spin for circularly polarized light. } 
	\label{fig:setup}
\end{figure}

Consider one-photon ionization of randomly oriented chiral molecules under isotropic illumination. To detect a net current, one needs to measure photoelectrons in all possible directions. Since the measurement, the system, and the light are completely isotropic, then there will be no net current in the system. However, if one introduces a spin detection axis, $\unitvec{s}$, then it breaks the symmetry and a net current along $\unitvec{s}$ is symmetry allowed (see Fig. \ref{fig:setup}). The current conditioned on spin can only have a component along the spin detection axis. Since the current is a time-odd polar vector while spin is a time-odd axial vector, then the projection of the current on the spin axis is a time-even pseudoscalar. Its pseudoscalar nature dictates that such current can only emerge in chiral molecules. As a time even quantity, it does not require breaking of the time-reversal symmetry of the Hamiltonian, therefore, the current is absent without spin conditioning and emerges as soon as the measurement is either spin-conditioned or postselected. 

To address the strength and nature of the these pseudoscalar correlations, we use our geometric formalism developed for spin-resolved one-photon photoionization of chiral molecules \cite{flores2024_2,flores2024_3}. It allows us to identify and quantify the molecular pseudoscalrs underlying both time-even  spin-current correlations as well as time-odd  triple correlations between the spin of the photoelectron, the photoelectron current, and the photon spin. We show that these correlations are enabled by two complementary geometric mechanisms that arise solely from electric dipole interactions and the geometric properties of the photoionization dipoles in real space and spin space. 

The first mechanism is mediated by the momentum-resolved Bloch pseudovector $\fullvec{\mathcal{S}}_{\fullvec{k}}$ defined below. It operates in fields of arbitrary polarization, including fully isotropic polarization. The second, time-odd, mechanism is mediated by the spin-resolved propensity field $\vec{\mathbb{B}}_{\fullvec{k}}$ which is a natural extension of 
its spin-averaged precursor \cite{ordonez2019propensity,ayuso2022} responsible for photoelectron circular dichroism \cite{ritchie1976theory,powis2000,bowering,nahon,janssen} and is
also defined below. This molecular property requires breaking time-reversal symmetry and can be activated by photon spin. To quantify our results we use synthetic chiral argon system employed in Ref. \cite{flores2024_3}, which are constructed by combining excited-state orbitals:
\begin{align}
	| \psi_{m,\mu}^{\pm} \rangle_p = \dfrac{1}{\sqrt{2}} \left( |4p_{m},\mu\rangle \pm |4d_{m},\mu\rangle \right).
	\label{eq:p_state}
\end{align}
\begin{align}
	| \psi_{m,\mu}^{\pm} \rangle_c = \dfrac{1}{\sqrt{2}} \left( |4p_{m},\mu\rangle \pm i |4d_{m},\mu\rangle \right).
	\label{eq:c_state}
\end{align}
These sates $| \psi_{m,\mu}^{\pm} \rangle_p$ and $| \psi_{m,\mu}^{\pm} \rangle_c$ are inspired by analogous chiral hydrogenic states \cite{ordonez2019propensity}. Unlike hydrogen, the multielectron core potential in argon breaks inversion symmetry: the \emph{synthetic chirality} is stabilized by electron correlations.

Now, using perturbation theory, the full spinor valued electron wave-function at the end of the ionizing pulse can be written as\footnote{Superscripts $L$ and $M$ are used to denote quantities in the laboratory and molecular frames, respectively. Vectors in the molecular frame, $\fullvec{a}^M$, are transformed into the laboratory frame using the relation $\fullvec{a}^L = R_\rho \fullvec{a}^M$, where $R_\rho$ is the Euler rotation matrix. Expressions without any superscript are to be understood as fully defined in the molecular frame.}
\begin{subequations} \label{eq:wavefunction}
	\begin{equation}
		|\psi\rangle=|\psi_{o}\rangle+ \sum_{I,\mu^M} \int d\Theta_k^M\, c_{I,\fullvec{k}^M,\mu^M}| I \Psi_{I,\fullvec{k}^M,\mu^M}^{(-)}\rangle,
		\label{eq:app:psi}
	\end{equation}
	\begin{equation}
		c_{I,\fullvec{k}^M,\mu^M} = i \left( \fullvec{D}_{I,\fullvec{k}^M,\mu^M}^L\cdot\fullvec{E}^L \right)
	\end{equation}
\end{subequations}
where $|\psi_{o}\rangle$ is the ground state of the molecule, $|I\Psi_{I,\fullvec{k}^M,\mu^M}^{(-)}\rangle$ is the fully spin-coupled continuum state with momentum $\fullvec{k}^M$ for the ionic channel $I$, $\mu=\pm\frac{1}{2}$ labels the photoelectron spin projection onto the molecular z-axis, $\fullvec{D}_{I,\fullvec{k}^M,\mu^M}^L=\langle \Psi_{I,\fullvec{k}^M,\mu^M}^{(-)} | \fullvec{d}^L | \psi_o \rangle$ is the spin-resolved transition dipole, and $\fullvec{E}^L$ describes the light field. The momentum and spin resolved photoionization rate $W^L(\unitvec{k}^L,\unitvec{s}^L,\rho)$ for a given orientation $\rho$ is obtained by projecting the full wavefunction\footnote{For brevity, we will drop the label $I$ in the succeeding expressions.}, Eq. \eqref{eq:wavefunction}, onto the scattering and ionic states with energy $\mathcal{E}$, then projecting the spin onto the axis $\unitvec{s}^L$:
\begin{align}
	\hat{\mathsf{P}}_{\mathcal{E}} =& \sum_{\mu,\nu} \int d\Theta_k |\Psi_{\fullvec{k},\mu}^{(-)} \rangle ( \hat{\mathsf{P}}_{\unitvec{s}} )_{\nu,\mu} \langle  \Psi_{\fullvec{k},\nu}^{(-)} | .
	\label{eq:projector}
\end{align}
Here $\hat{\mathsf{P}}_{\unitvec{s}}=(\mathbb{I} + \unitvec{s} \cdot \unitvec{\sigma})/2$ is a spin projection operator with respect to $\unitvec{s}$, and $\unitvec{\sigma}$ is the vector of Pauli spin matrices. Performing the necessary operations, we obtain 
\begin{align}
	W^L&(\unitvec{k}^L,\unitvec{s}^L,\rho) \nonumber \\
	=& \dfrac{1}{2} \sum_{I,\mu,\nu} \int d\rho \left(  \fullvec{D}_{I,\fullvec{k}^M,\mu}^{L*}\cdot\fullvec{E}^{L*} \right) \left(  \fullvec{D}_{I,\fullvec{k}^M,\nu}^{L}\cdot\fullvec{E}^{L} \right) \nonumber \\
	&\times \left( \delta_{\mu,\nu} + \unitvec{s}^L\cdot\unitvec{\sigma}_{\nu,\mu}^L\right). \label{eq:yield}  
\end{align}
The photoelectron current conditioned on the spin detection axis $\unitvec{s}^L$ can then be calculated as follows: 
\begin{align}
	\fullvec{j}^L(\unitvec{s}^L) =& \dfrac{ \int d\rho \int d\Theta_k^L  W^L(\unitvec{k}^L,\unitvec{s}^L,\rho)\fullvec{k}^L }{ \int d\rho \int d\Theta_s^L\int d\Theta_k^L  W^L(\unitvec{k}^L,\unitvec{s}^L,\rho) } 
	\label{eq:currents-total}
\end{align} 
where $\int d\rho$, $\int d\Theta_k^L$, and $\int d\Theta_s^L$ denote averaging over all molecular orientations, photoelectron momenta, and spin detection axes, respectively (see Supplementary Material for details).  

\begin{figure*}[htb!]
	\centering
	\includegraphics[width=0.95\textwidth]{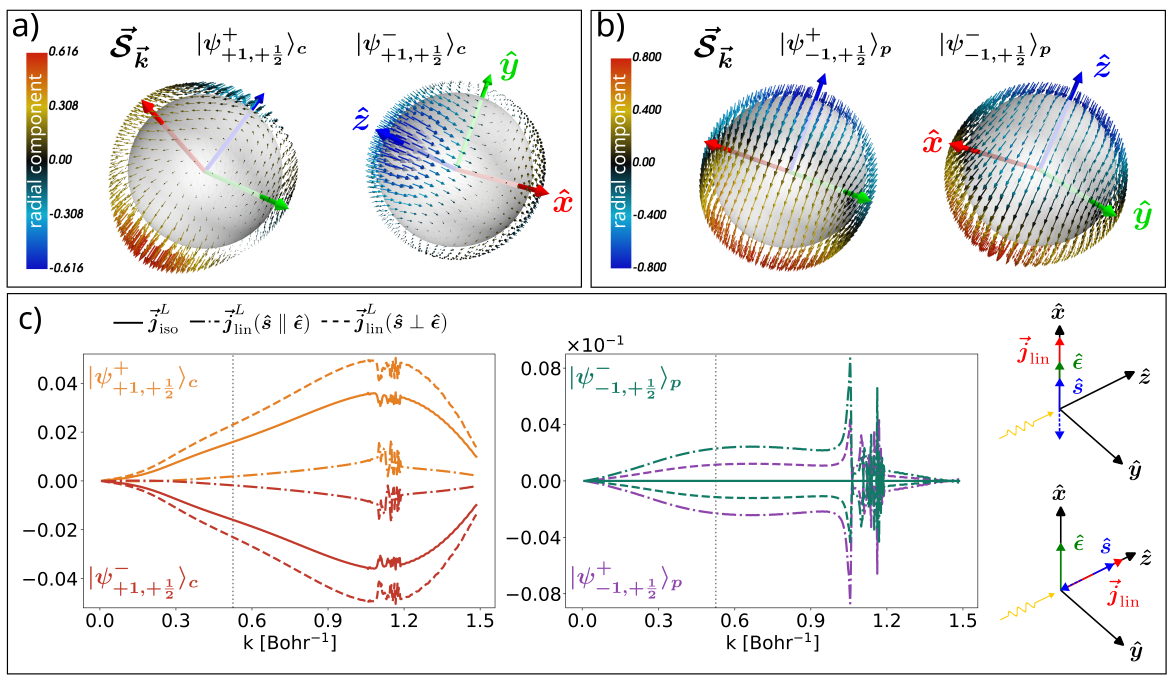}
	\caption{(a,b) The momentum-resolved Bloch vector $\fullvec{\mathcal{S}}_{\fullvec{k}}$ in k-space for the chiral states $|\psi_{1,\frac{1}{2}}^\pm\rangle_c$ and $|\psi_{-1,\frac{1}{2}}^\pm\rangle$. The spheres correspond to  $\fullvec{\mathcal{S}}_{\fullvec{k}}$ with $|\fullvec{k}|=0.53\,\text{Bohr}^{-1}$ where the vectors are scaled as $|\fullvec{\mathcal{S}}_{\fullvec{k}}|$, and colored according to $(\fullvec{\mathcal{S}}_{\fullvec{k}}\cdot\unitvec{k})$. (c) Comparison of the photoelectron current $\fullvec{j}_{\text{iso}}$ under isotropic illumination (solid line) and that of linearly polarized light with collinear $\fullvec{j}_{\text{lin}}(\unitvec{s}\parallel\unitvec{\epsilon})$ [dotted-dashed line] and orthogonal $\fullvec{j}_{\text{lin}}(\unitvec{s}\perp\unitvec{\epsilon})$ [dashed line] detection geometries. The gray line marks $|\fullvec{k}|=0.53\,\text{Bohr}^{-1}$ corresponding to (a,b). The rapidly oscillating behavior of $\fullvec{j}$ at higher values of \(k\) are due to the Fano resonances, leading up to the ionization threshold for the 3s electrons \cite{Samson2002,Carlstroem2024spinpolspectral}. }
	\label{fig:texture-Bloch}
\end{figure*}

Let us first consider randomly oriented chiral molecules under isotropic illumination by linearly polarized light: 
\begin{align}
	\fullvec{E}^L =& E_\omega^L \left( \sin\theta_p \cos\varphi_p \unitvec{x}^L + \sin\theta_p \sin\varphi_p \unitvec{y}^L + \cos\theta_p \unitvec{z}^L \right), 
	\label{eq:random_E}
\end{align}
where  the angles $(\theta_p$, $\varphi_p)$ describe all possible polarization directions. Using Eqs. \eqref{eq:yield}-\eqref{eq:random_E}, and averaging over all polarization directions $\int d\Theta_p$, the resulting photoelectron current is 
\begin{align}
	\fullvec{j}_{\text{iso}}^{L}  =& \dfrac{1}{3 S_0} \left( \dfrac{1}{k} \int d\fullvec{\Theta}_k^M \cdot \fullvec{\mathcal{S}}_{\fullvec{k}}^M  \right) \unitvec{s}^L,
	\label{eq:jS_rand}
\end{align}
where 
\begin{align}
	S_0 = \int d\Theta_k \left( \left|\fullvec{D}_{\fullvec{k},\frac{1}{2}}\right|^2+ \left|\fullvec{D}_{\fullvec{k},-\frac{1}{2}}\right|^2 \right),
\end{align}
is proportional to the total 
ionization yield. Equation \eqref{eq:jS_rand} shows that the current is proportional to the flux of the vector $\fullvec{\mathcal{S}}_{\fullvec{k}}^M$ through the surface of the energy shell ($d\fullvec{\Theta}_k^M=d\Theta_{k}^M\unitvec{k}^Mk^2$). Mathematically, $\fullvec{\mathcal{S}}_{\fullvec{k}}=\text{Tr}\left( \tilde{\varrho} \, \unitvec{\sigma} \right) $ is a Bloch pseudovector that describes spin orientation in the degenerate two-level system formed by spin-up and spin-down photoelectron continuum states. It incorporates both populations and coherences within this system. The reduced density matrix, $\tilde{\varrho}=\fullvec{D}_{\fullvec{k},\mu}^* \cdot \fullvec{D}_{\fullvec{k},\nu}$, emerges after performing a trace over the degenerate ionization channels of the full density matrix, then averaging over all molecular orientations \cite{flores2024_3}. Moreover, the Bloch vector $\fullvec{\mathcal{S}}_{\fullvec{k}}$ defines a spin-texture on the energy shell for a chiral molecule fixed in space and photoionized by isotropic illumination, see Fig. \ref{fig:texture-Bloch}(a,b). 

\begin{figure*}[t!]
	\centering
	\includegraphics[width=0.8\textwidth]{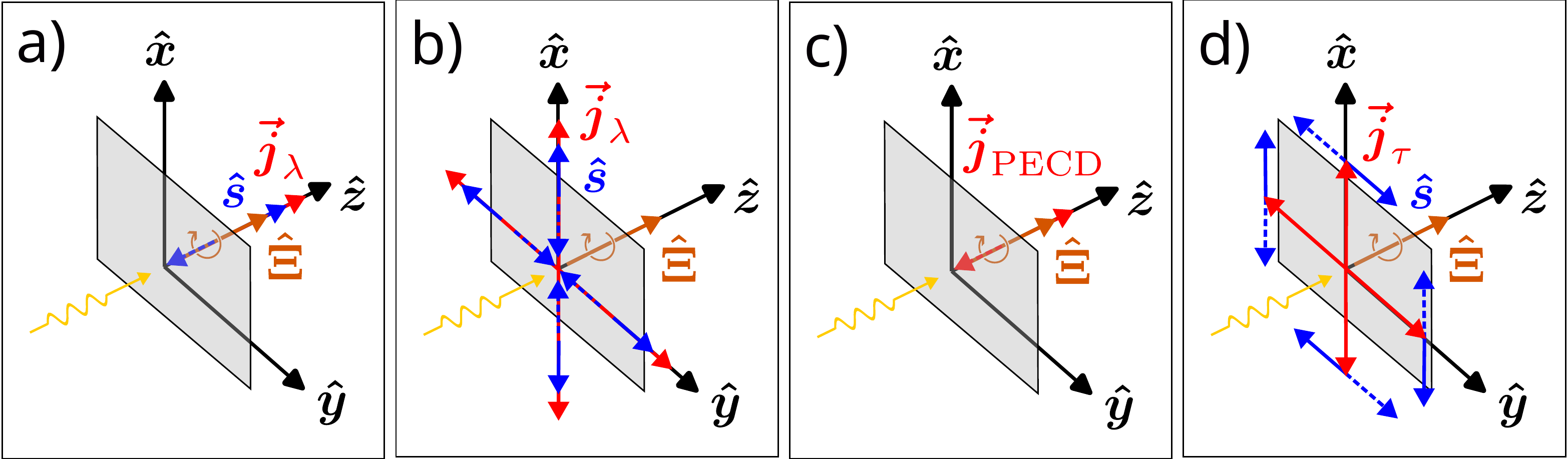}
	\caption{Schematic of the resulting photoelectron currents for circularly polarized light in the laboratory frame. The solid and dashed blue vectors denote the opposite directions of spin for opposite enantiomers. (a,b) The current $\fullvec{j}_r$ is enabled by the Bloch vector $\fullvec{\mathcal{S}}_{\fullvec{k}}$ which results in a collinear locking of the photoelectron current and its spin $\fullvec{j}\parallel\unitvec{s}$. (c,d) The currents $\fullvec{j}_{\text{PECD}}$ and $\fullvec{j}_\tau$ are enabled by the spin-resolved propensity field $\vec{\mathbb{B}}_{\fullvec{k}}$. The PECD $\fullvec{j}_{\text{PECD}}$ current is not spin-sensitive, while $\fullvec{j}_\tau$ presents a triple locking of the photoelectron current, its spin with the photon spin in three mutually orthogonal directions.}
	\label{fig:scheme-circ}
\end{figure*}

The current $\fullvec{j}_{\text{iso}}^{L}$, Eq. \eqref{eq:jS_rand}, shows that photoelectrons with opposite spin projections along $\unitvec{s}^L$ are correlated to opposite enantiomers, i.e., 
\begin{equation}
	\fullvec{j}_{\text{iso}}^{L(S)}(\unitvec{s}^L) = \fullvec{j}_{\text{iso}}^{L(R)}(-\unitvec{s}^L),
\end{equation}
and that the current is collinearly `locked' to the direction of the spin-detection axis $\unitvec{s}^L$: the net photoelectron current is along the spin detection axis.
The strength of these correlations is defined by Eq. \eqref{eq:jS_rand}: the pseudoscalar describing spin-current correlations is a flux (or the net radial component) of the momentum-resolved Bloch vector. This expression shows the link between time-odd spin- molecular orientation correlations \cite{flores2024_3} driven by  the projection of the net Bloch vector, $\fullvec{\mathcal{S}}_{\text{net}}=\int d\Theta_k \fullvec{\mathcal{S}}_{\fullvec{k}}$, on to a molecular bond  and its time-even spin-current correlations driven by the net radial component (or the flux of the momentum resolved Bloch vector through the energy shell). This result establishes the momentum-resolved Bloch vector as a primary molecular property defining all spin-dependent pseudoscalar correlations in photoionization of chiral molecules.

Next we discuss additional biases that may be imparted by light with well defined polarization (linear or circular) and propagation direction. Repeating the same process for linearly polarized $\fullvec{E}^L = E_\omega^L \unitvec{\epsilon}^L$, we obtain the photoelectron current 
\begin{subequations} \label{eq:J-lin}
	\begin{align}
		\fullvec{j}_{\text{lin}}^{L} =& \dfrac{1}{5S_0}\left[ \dfrac{1}{k} \int d\fullvec{\Theta}_k^M \cdot \left( 2\fullvec{\mathcal{S}}_{\fullvec{k}}^M + \fullvec{\mathcal{S'}}_{\fullvec{k}}^M \right) \right] \unitvec{s}^L \nonumber \\
		&-\dfrac{1}{5S_0}\left[ \dfrac{1}{k} \int d\fullvec{\Theta}_k^M \cdot \left( \fullvec{\mathcal{S}}_{\fullvec{k}}^M + 3 \fullvec{\mathcal{S'}}_{\fullvec{k}}^M \right) \right] \left( \unitvec{s}^L \cdot \unitvec{\epsilon}^L\right) \unitvec{\epsilon}^L
		\label{eq:j-lin}
	\end{align}
	\begin{align}
		\fullvec{\mathcal{S'}}_{\fullvec{k}}^M =& \text{Re}\left[ \sum_{\mu,\nu} \left( \fullvec{D}_{\fullvec{k}^M,\mu}^{M*} \cdot \unitvec{\sigma}_{\nu,\mu}^M  \right) \fullvec{D}_{\fullvec{k}^M,\nu}^M \right].
	\end{align}
\end{subequations}
The additional vector $\fullvec{\mathcal{S'}}_{\fullvec{k}}^M$ is the directional bias introduced by the well-defined direction of light polarization. Hence, the photoelectron current now arises as the flux of an effective Bloch pseudovector $\fullvec{\mathsf{S}}_{\fullvec{k}} = a \fullvec{\mathcal{S}}_{\fullvec{k}}^M + b \fullvec{\mathcal{S'}}_{\fullvec{k}}^M$ through the surface of the energy shell. The current $\fullvec{j}_{\text{lin}}^{L}$ Eq. \eqref{eq:J-lin} can be measured using two possible detection geometries: (i) collinear $\unitvec{s}\parallel\unitvec{\epsilon}$ and (ii) orthogonal $\unitvec{s}\perp\unitvec{\epsilon}$, see Fig. \ref{fig:texture-Bloch}c. Enhancement or reduction of $\fullvec{j}_{\text{lin}}^{L}$ with respect to $\fullvec{j}^L_{\text{iso}}$ is then controlled by $\fullvec{\mathcal{S'}}_{\fullvec{k}}^M$. The comparison of the currents $\fullvec{j}_{\text{iso}}^{L}$ and $\fullvec{j}_{\text{lin}}^{L}$ is shown in Fig. \ref{fig:texture-Bloch}c. It can be seen that the changes in the current are sensitive to both the chiral state and detection geometry.

Now, consider circularly polarized light $\fullvec{E}^{L}=E_\omega^L(\unitvec{x}^L + i \xi \unitvec{y}^L)/\sqrt{2}$, where, $\xi=\pm1$ is a dichroic parameter characterizing the direction of rotation of the light polarization vector. The resulting current will have the following components (Fig.\ref{fig:scheme-circ}): 
\begin{align}
	\fullvec{j}_{\text{circ}}^L =  \fullvec{j}_{\text{PECD}}^L + \fullvec{j}_{\tau}^L + \fullvec{j}_{\lambda}^L.
    \label{eq:currentj_circ}
\end{align}
The terms $\fullvec{j}_{\text{PECD}}^L$ and $\fullvec{j}_\tau^L$ are enabled by the pseudovector, $(\vec{\mathbb{B}}_{\fullvec{k}})_{\mu,\nu} =  i \fullvec{D}_{\fullvec{k},\mu}^* \times \fullvec{D}_{\fullvec{k},\nu}$, which we refer to as the spin-resolved propensity field since it encodes the propensity rules for ionization. Explicitly, we have 
\begin{align}
    \fullvec{j}_{\text{PECD}}^L 
    =& \dfrac{1}{2 S_0} \left(  \dfrac{1}{k} \int d \fullvec{\Theta}_k^M \cdot \fullvec{\mathcal{B}}_{\fullvec{k}^M}^M\right) \unitvec{\Xi}^L,
    \label{eq:PECD}
\end{align}
where, $\fullvec{\mathcal{B}}_{\fullvec{k}^M}^M = \text{Tr}( \vec{\mathbb{B}}^M_{\fullvec{k}})$, and yields the familiar  PECD (photoelectron circular dichroism) current, and not spin sensitive, thereby presenting a ``background'' in spin-conditioned measurements\footnote{Equation \eqref{eq:PECD} is equal to the PECD current we derived in \cite{ordonez2018generalized}, to the original expression derived by Ritchie \cite{ritchie1976theory}, and to the coefficient $D$ introduced by Cherepkov \cite{cherepkov1979spin,cherepkov1981theory,cherepkov1983manifestations,flores2024_2}.}.

\begin{figure*}[t!]
	\centering
	\includegraphics[width=0.95\textwidth]{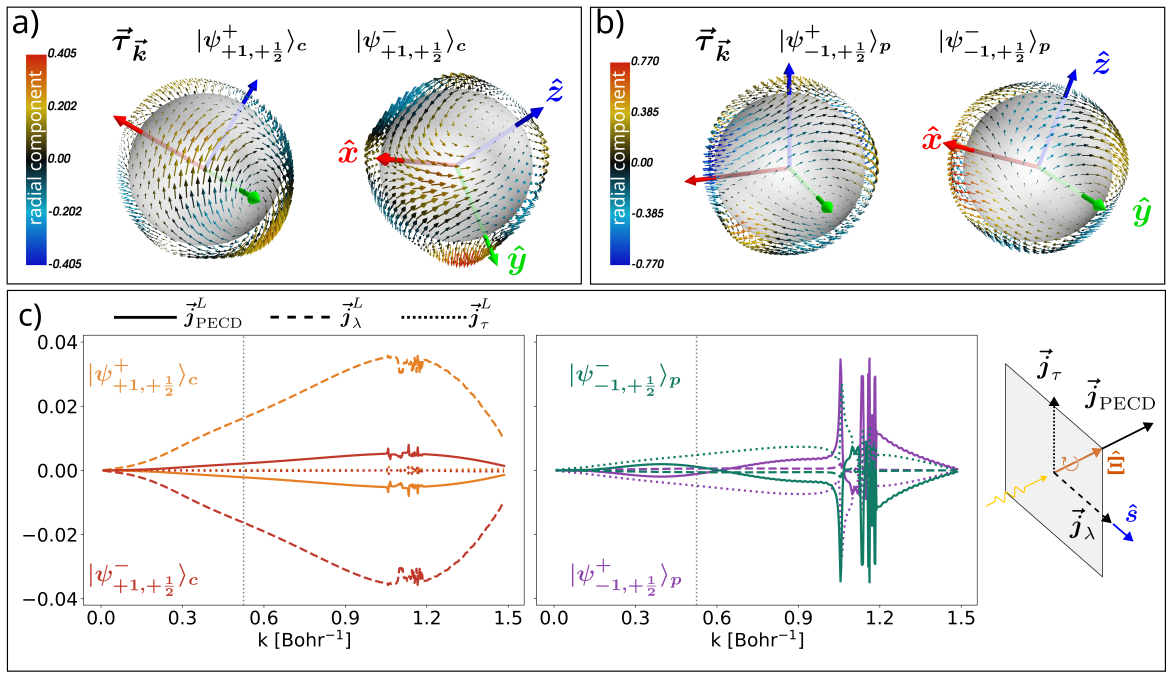}
	\caption{(a,b) The torque vector $\fullvec{\tau}_{\fullvec{k}}$ in k-space for the chiral states $|\psi_{1,\frac{1}{2}}^\pm\rangle_c$ and $|\psi_{-1,\frac{1}{2}}^\pm\rangle$. The spheres correspond to  $\fullvec{\tau}_{\fullvec{k}}$ with $|\fullvec{k}|=0.53\,\text{Bohr}^{-1}$ where the vectors are scaled as $|\fullvec{\tau}_{\fullvec{k}}|$, and colored according to $(\fullvec{\tau}_{\fullvec{k}}\cdot\unitvec{k})$. (c) Comparison of the components of the photoelectron current $\fullvec{j}_{\text{circ}}^L$, Eq. \eqref{eq:currentj_circ}, under circularly polarized light where the $\unitvec{s}\perp\unitvec{\xi}$. 
    }
	\label{fig:texture-tau}
\end{figure*}

The term $\fullvec{j}_\tau^L$ presents a transversal spin polarization:
\begin{align}
		\fullvec{j}_{\tau}^L 
		=& \dfrac{1}{4 S_0} \left( \dfrac{1}{k} \int d \fullvec{\Theta}_k^M \cdot \fullvec{\tau}^M\right) \left( \unitvec{s}^L \times  \unitvec{\Xi}^L \right)
		\label{eq:vort}
\end{align}
which is proportional to the flux of the spin torque vector $\fullvec{\tau}_{\fullvec{k}^M}^M=\text{Tr}(\unitvec{\sigma}^M\times\vec{\mathbb{B}}_{\fullvec{k}^M}^M)$. This current is maximal when the photoelectron spin detection axis $\unitvec{s}^L$ is orthogonal to the photon spin $\unitvec{\Xi}^L=(-i\fullvec{E}^{L*}\times\fullvec{E}^L)/|\fullvec{E}^L|^2=\xi\unitvec{z}^L$. Thus, it is confined in the polarization plane and the respective spin-current correlations are transversal\footnote{Equation \eqref{eq:vort} is equal to the coefficient $C$ introduced by Cherepkov \cite{cherepkov1979spin,cherepkov1981theory,cherepkov1983manifestations,flores2024_2}.}, i.e., opposite enantiomers produce opposite spin vortices, such that the spin polarization direction depends on the photoelectron momentum in a way reminiscent of the Rashba effect in solids. The momentum-resolved spin-torque vector $\fullvec{\tau}_{\fullvec{k}^M}^M$, see Fig. \ref{fig:texture-tau}(a,b), quantifies the intrinsic coupling between the photon and photoelectron spin mediated by the spin-resolved propensity field, and defines the direction of the torque exerted on the photoelectron spin by the photon spin for every direction of the photoelectron momentum in the molecular frame. 

The term $\fullvec{j}_\lambda^L$ is enabled by an effective Bloch pseudovector $\fullvec{\mathsf{S}}_{\fullvec{k}}$ and has a similar form as that of Eq. \eqref{eq:J-lin}:
\begin{align}
	\fullvec{j}_\lambda^L =& \dfrac{1}{10S_0} \left[ \dfrac{1}{k} \int d\fullvec{\Theta}_k^M \cdot \left(3\fullvec{\mathcal{S}}_{\fullvec{k}}^M + \fullvec{\mathcal{S'}}_{\fullvec{k}}^M\right) \right] \unitvec{s}^L\nonumber \\
	&+\dfrac{1}{10S_0} \left[ \dfrac{1}{k} \int d\fullvec{\Theta}_k^M \cdot \left(\fullvec{\mathcal{S}}_{\fullvec{k}}^M -3 \fullvec{\mathcal{S'}}_{\fullvec{k}}^M\right) \right] (\unitvec{s}^L\cdot\unitvec{\Xi}^L)\unitvec{\Xi}^L,
	\label{eq:sink}
\end{align}
which is enantio-sensitive but not dichroic\footnote{This is contained in the coefficients $B_1$ and $B_2$ introduced by Cherepkov \cite{cherepkov1979spin,cherepkov1981theory,cherepkov1983manifestations,flores2024_2}}. The collinear locking between the photoelectron current and spin is still present and indicates a longitudinal spin polarization which can be measured either via collinear or orthogonal detection geometry with respect to $\unitvec{\Xi}^L$. 

The three components contributing to  $\fullvec{j}_{\text{circ}}^L$, Eq. \eqref{eq:currentj_circ}, become mutually orthogonal when $\unitvec{s}^L\perp\unitvec{\Xi}^L$. In this case they can be measured independently, and a comparison of each term in $\fullvec{j}_{\text{circ}}^L$ is shown in Fig. \ref{fig:texture-tau}(c). It can be seen that the components are sensitive to the chiral state, however, both $\fullvec{j}_\tau^L$ and $\fullvec{j}_\lambda^L$ can be significantly larger than $\fullvec{j}_{\text{PECD}}^L$, i.e., $|\fullvec{j}_\lambda^L|/|\fullvec{j}_{\text{PECD}}^L|\lesssim8$ for the state $|\psi_{1,\frac{1}{2}}^\pm\rangle_c$ while  $|\fullvec{j}_\tau^L|/|\fullvec{j}_{\text{PECD}}^L|\lesssim2$ for the state $|\psi_{-1,\frac{1}{2}}^\pm\rangle_p$. 

We  have established two fundamental mechanisms of CISS in photoionization operating in time-even  and time-odd (i.e. involving photon spin) set-ups: (i) the momentum-resolved Bloch pseudovector $\fullvec{\mathcal{S}}_{\fullvec{k}}$ which enables collinear locking of the photoelectron current with the photoelectron spin and operates in arbitrary light fields, even under isotropic illumination, and (ii) the spin-resolved propensity field $\vec{\mathbb{B}}_{\fullvec{k}}$ which enables orthogonal triple locking between the photoelectron spin, its current and the photon spin. We have verified that the spin-conditioned photoelectron currents derived in Eqs. \eqref{eq:j-lin}, \eqref{eq:sink}, and \eqref{eq:vort} are consistent \cite{flores2024_2} with the kinematic predictions of Cherepkov \cite{cherepkov1979spin,cherepkov1981theory,cherepkov1983manifestations}. Our approach reveals the dynamical origin of these phenomena. Importantly, it also leads to a qualitatively new prediction: the generation of a spin-conditioned current even for randomly oriented chiral molecules under isotropic illumination.

O.S., A.F.O. and P.C.F. acknowledge ERC-2021-AdG project ULISSES, grant agreement No 101054696. Views and opinions expressed are however those of the author(s) only and do not necessarily reflect those of the European Union or the European Research Council. Neither the European Union nor the granting authority can be held responsible for them. A.F.O. acknowledges funding from the Royal Society URF/R1/201333, URF/ERE/210358, and URF/ERE/231177 and from the Deutsche Forschungsgemeinschaft (DFG, German Research Foundation) - 543760364.

\begin{widetext}
\appendix
\section*{Derivation of the Photoelectron Current for Isotropic Illumination}
The photoelectron current conditioned on the spin detection axis $\unitvec{s}^L$ is calculated as follows: 
\begin{subequations}
\begin{align}
	\fullvec{j}^L(\unitvec{s}^L) =& \dfrac{ \int d\rho \int d\Theta_k^L  W^L(\unitvec{k}^L,\unitvec{s}^L,\rho)\fullvec{k}^L }{ \int d\rho \int d\Theta_s^L\int d\Theta_k^L  W^L(\unitvec{k}^L,\unitvec{s}^L,\rho) } 
	\label{eq-app:currents-total}
\end{align} 
\begin{align}
	W^L(\unitvec{k}^L,\unitvec{s}^L,\rho) =& \dfrac{1}{2} \sum_{\mu,\nu} \int d\rho \left(  \fullvec{D}_{\fullvec{k}^M,\mu}^{L*}\cdot\fullvec{E}^{L*} \right) \left(  \fullvec{D}_{\fullvec{k}^M,\nu}^{L}\cdot\fullvec{E}^{L} \right)\left( \delta_{\mu,\nu} + \unitvec{s}^L\cdot\unitvec{\sigma}_{\nu,\mu}^L\right),
	\label{eq-app:yield}
\end{align}
\end{subequations}
wherein, $\int d\rho$, $\int d\Theta_s^L$, and $\int d\Theta_k^L$ denotes averaging over all molecular orientations, photoelectron momentum, and spin detection axis, respectively. The vectors that appear on the right-hand side of Eq. \eqref{eq-app:yield} can be grouped into two sets: (i) vectors that are fixed in the molecular frame such as the dipole transition vectors $\fullvec{D}_{I,\fullvec{k}^M,\mu^M}^M$, photoelectron momentum $\fullvec{k}^M$, and photoelectron spin quantization axis $\unitvec{\sigma}_{\mu_1^M,\mu_2^M}^M$, and (ii) vectors that are fixed in the laboratory frame such as spin detection axis $\unitvec{s}^L$ and the electric field $\fullvec{E}^L$. This will then allow us to use the technique in Ref. \cite{andrews1977three} in evaluating the orientation averaging $\int d\rho$ such that the resulting quantity can be expressed as $\sum_{ij}g_i M_{ij} f_j$, where, $g_i$ and $f_i$ are rotational invariants that are constructed from the two sets of vectors and $M_{ij}$ is the coupling between the two rotational invariants. For our purposes, the following vector identities will be relevant: 
\begin{align}
	\int d\rho (\fullvec{a}^L\cdot\fullvec{u}^L)  \fullvec{b}^L = \dfrac{1}{3} (\fullvec{a}^M\cdot\fullvec{b}^M) \fullvec{u}^L
	\label{eq:rank2}
\end{align}
\begin{align}
	\int d\rho (\fullvec{a}^L\cdot\fullvec{u}^L)(\fullvec{b}^L\cdot\fullvec{v}^L) \fullvec{c}^L = \dfrac{1}{6} [(\fullvec{a}^M \times \fullvec{b}^M) \cdot \fullvec{c}^M] (\fullvec{u}^L \times \fullvec{v}^L)
	\label{eq:rank3}
\end{align}
\begin{align}
	\int d\rho (\fullvec{a}^L\cdot\fullvec{u}^L)(\fullvec{b}^L\cdot\fullvec{v}^L)(\fullvec{c}^L\cdot\fullvec{w}^L)\fullvec{d}^L = \dfrac{1}{30}
	\begin{bmatrix}
		(\fullvec{a}^M\cdot\fullvec{b}^M)(\fullvec{c}^M\cdot\fullvec{d}^M) \\
		(\fullvec{a}^M\cdot\fullvec{c}^M)(\fullvec{b}^M\cdot\fullvec{d}^M) \\
		(\fullvec{a}^M\cdot\fullvec{d}^M)(\fullvec{b}^M\cdot\fullvec{c}^M)
	\end{bmatrix}^T
	\begin{bmatrix}
		4   &   -1  &   -1  \\
		-1  &   4   &   -1  \\
		-1  &   -1  &   4
	\end{bmatrix}
	\begin{bmatrix}
		(\fullvec{u}^L\cdot\fullvec{v}^L)\fullvec{w}^L \\
		(\fullvec{u}^L\cdot\fullvec{w}^L)\fullvec{v}^L \\
		(\fullvec{v}^L\cdot\fullvec{w}^L)\fullvec{u}^L
	\end{bmatrix}
	\label{eq:rank4}
\end{align}

Using Eqs. \eqref{eq:rank2}-\eqref{eq:rank3}, we can easily evaluate the normalization factor as follows: 
\begin{align}
\int d\rho & \int d\Theta_s^L\int d\Theta_k^L  W^L(\unitvec{k}^L,\unitvec{s}^L,\rho) \nonumber \\
=& \int d\Theta_s^L\int d\Theta_k^M  \int d\rho  W^L(\unitvec{k}^M,\unitvec{s}^M,\rho) \nonumber \\
=& \dfrac{1}{2} \int d\Theta_s^L\int d\Theta_k^M  \sum_{\mu,\nu} \int d\rho \left(  \fullvec{D}_{\fullvec{k}^M,\mu}^{L*}\cdot\fullvec{E}^{L*} \right) \left(  \fullvec{D}_{\fullvec{k}^M,\nu}^{L}\cdot\fullvec{E}^{L} \right)\left( \delta_{\mu,\nu} + \unitvec{s}^L\cdot\unitvec{\sigma}_{\nu,\mu}^L\right) \nonumber \\
=& \dfrac{1}{6} \int d\Theta_s^L\int d\Theta_k^M  \sum_{\mu,\nu} \left[\left(  \fullvec{D}_{\fullvec{k}^M,\mu}^{M*}\cdot\fullvec{D}_{\fullvec{k}^M,\nu}^{M} \right) |\fullvec{E}^{L}|^2 \delta_{\mu,\nu}  \right] \nonumber \\
&+ \dfrac{1}{12} \left\{ \int d\Theta_k^M  \sum_{\mu,\nu} \left[ \unitvec{\sigma}_{\nu,\mu}^M \cdot \left( \fullvec{D}_{\fullvec{k}^M,\mu}^{M*}\times\fullvec{D}_{\fullvec{k}^M,\nu}^{M} \right) \right] \right\} \left[ \left( \int d\Theta_s^L \unitvec{s}^L  \right)\cdot \left( \fullvec{E}^{L*} \times \fullvec{E}^L \right) \right] \nonumber \\
=& \dfrac{1}{6} \left( \sum_{\mu} \int d\Theta_k^M \left| \fullvec{D}_{\fullvec{k}^M,\mu}^{M} \right|^2 \right) |\fullvec{E}^{L}|^2
\end{align}
The second line follows from the definition of a rotated function $W^L(\unitvec{k}^L,\unitvec{s}^L,\rho)=W^M(\unitvec{k}^M,\unitvec{s}^M,\rho)$ \cite{brink1968angular}, and we have interchanged the order of integration to perform change of variable $\unitvec{k}^M=R_\rho^{-1}\unitvec{k}^L$ and again interchanged the order of integration since $\unitvec{k}^M$ is now an integration variable independent of $\rho$. Meanwhile, the last line follows from the vanishing of $\int d\Theta_s^L \unitvec{s}^L$. 

Let us now consider the spin-conditioned photoelectron current under isotropic illumination: 
\begin{equation}
\fullvec{j}_{\text{iso}}^L = \dfrac{ \int d\Theta_p \int d\rho \int d\Theta_k^L  W^L(\unitvec{k}^L,\unitvec{s}^L,\rho)\fullvec{k}^L }{ \int d\Theta_p \int d\rho \int d\Theta_s^L\int d\Theta_k^L  W^L(\unitvec{k}^L,\unitvec{s}^L,\rho) } ,
\label{eq-app:J_iso}
\end{equation}
where, $\int d\Theta_p$ denotes averaging over all orientations of the light field: 
\begin{align}
    \fullvec{E}^L_p = E_\omega^L \left( \sin\theta_p\cos\varphi_p \unitvec{x}^L + \sin\theta_p\sin\varphi_p \unitvec{y}^L + \cos\theta_p \unitvec{z}^L \right)
    \label{eq:app_E_iso}
\end{align}
The relevant quantity to calculate is the the numerator of Eq. \eqref{eq-app:J_iso}, i.e., 
\begin{align}
\int d\Theta_k^M & \int d \Theta_p \int d\rho W^M(\unitvec{k}^M,\unitvec{s}^M,\rho) \fullvec{k}^L \nonumber \\
=& \dfrac{1}{2} \sum_{\mu_1^M,\mu_2^M} \int d\Theta_k^M \int d \Theta_p \int d\rho  \left( \fullvec{D}_{\fullvec{k}^M,\mu}^{L*} \cdot\fullvec{E}_p^{L*} \right)   \left( \fullvec{D}_{\fullvec{k}^M,\nu}^{L} \cdot\fullvec{E}_p^{L} \right) \left( \delta_{\mu, \nu} + \unitvec{s}^L \cdot \unitvec{\sigma}_{\nu,\mu}\right) \fullvec{k}^L
\label{eq-app:J_iso0}
\end{align}
Using Eq. \eqref{eq:rank3}, the first term of Eq. \eqref{eq-app:J_iso0} simplifies into 
\begin{align}
\dfrac{1}{2}  \sum_{\mu} & \int d\Theta_k^M \int d \Theta_p \int d\rho \left| \fullvec{D}_{\fullvec{k}^M,\mu}^L\cdot\fullvec{E}_p^L \right|^2  \fullvec{k}^L \nonumber \\
=&
\dfrac{1}{12} \left\{ \left[ \sum_{\mu} \int d\Theta_k^M \left( \fullvec{D}_{\fullvec{k}^M,\mu}^{M*} \times \fullvec{D}_{\fullvec{k}^M,\mu}^{M} \right)  \right] \cdot \fullvec{k}^M \right\} \left[ \int d\Theta_p \left(\fullvec{E}^{L*}_p\times \fullvec{E}^L_p\right) \right] =0,
\label{eq:ave_Vec_term1}
\end{align}
which vanishes after averaging over all orientations of the field. Similarly, it follows from Eq. \eqref{eq:rank4} that the second term of Eq. \eqref{eq-app:J_iso} is now
\begin{align}
\dfrac{1}{2} \sum_{\mu,\nu} \int d\Theta_k^M & \int d\Theta_p \int d\rho  \left( \fullvec{D}_{\fullvec{k}^M,\mu}^{L*}\cdot\fullvec{E}_p^{L*} \right) \left( \fullvec{D}_{I,\fullvec{k}^M,\nu}^L\cdot\fullvec{E}_p^L \right)  \left( \unitvec{s}^L\cdot\unitvec{\sigma}_{\nu,\mu}^L\right) \fullvec{k}^L \nonumber \\
=\dfrac{1}{60} &
\begin{bmatrix}
    \sum\int d\Theta_k^M  \left( \fullvec{D}_{\fullvec{k}^M,\mu}^{M*}\cdot\fullvec{D}_{\fullvec{k}^M,\nu}^{M} \right) \left( \unitvec{\sigma}_{\nu,\mu}^M \cdot \fullvec{k}^M\right) \\
    \sum \int d\Theta_k^M \left( \fullvec{D}_{\fullvec{k}^M,\mu}^{M*}\cdot \unitvec{\sigma}_{\nu,\mu}^M  \right) \left( \fullvec{D}_{\fullvec{k}^M,\nu}^{M} \cdot \fullvec{k}^M\right) \\
    \sum \int d\Theta_k^M  \left( \fullvec{D}_{\fullvec{k}^M,\mu}^{M*}\cdot \fullvec{k}^M \right) \left( \fullvec{D}_{\fullvec{k}^M,\nu}^{M} \cdot \unitvec{\sigma}_{\nu,\mu}^M\right)
\end{bmatrix}^T 
\begin{bmatrix}
    4 & -1 & -1 \\
    -1 & 4 & -1 \\
    -1 & -1 & 4
\end{bmatrix}
\begin{bmatrix}
    \int d\Theta_p|\fullvec{E}_p^L|^2 \unitvec{s}^L \\
    \int d\Theta_p(\fullvec{E}_p^{L*}\cdot\unitvec{s}^L)\fullvec{E}_p^L \\
    \int d\Theta_p(\fullvec{E}_p^{L}\cdot\unitvec{s}^L)\fullvec{E}_p^{L*}
\end{bmatrix} \nonumber \\
= \dfrac{1}{60} & 
\begin{bmatrix}
    g_1 \\
    g_2 \\
    g_2^*
\end{bmatrix}^T
\begin{bmatrix}
    4 & -1 & -1 \\
    -1 & 4 & -1 \\
    -1 & -1 & 4
\end{bmatrix}
\begin{bmatrix}
    f_1 \\
    f_2 \\
    f_2^*
\end{bmatrix} \nonumber \\
=\dfrac{1}{30} & \left(2f_1 - \text{Re}[f_2]\right)g_1 - \dfrac{1}{30} \left(f_1 -3 \text{Re}[f_2]\right) \text{Re}[g_2] - \dfrac{1}{6} \text{Im}[f_2]\text{Im}[g_2]
\end{align}
Averaging over all orientations of the field we get:
\begin{align}
\dfrac{1}{2} \sum_{\mu,\nu} \int d\Theta_k^M & \int d\Theta_p \int d\rho  \left( \fullvec{D}_{\fullvec{k}^M,\mu}^{L*}\cdot\fullvec{E}_p^{L*} \right) \left( \fullvec{D}_{I,\fullvec{k}^M,\nu}^L\cdot\fullvec{E}_p^L \right)  \left( \unitvec{s}^L\cdot\unitvec{\sigma}_{\nu,\mu}^L\right) \fullvec{k}^L  \nonumber \\
=& \dfrac{1}{18} |\fullvec{E}^L|^2 \left\{ \int d\Theta_k^M  \left[ \sum_{\mu,\nu} \left( \fullvec{D}_{I,\fullvec{k}^M,\mu}^{M*}\cdot  \fullvec{D}_{I,\fullvec{k}^M,\nu}^{M} \right)  \unitvec{\sigma}_{\nu,\mu}^M  \right] \cdot \fullvec{k}^M \right\} \unitvec{s}^L \nonumber \\
=& \dfrac{1}{18} |\fullvec{E}^L|^2 \left[ \int d\Theta_k^M \left( \fullvec{\mathcal{S}}_{\fullvec{k}}^M \cdot \fullvec{k}^M \right) \right]
\unitvec{s}^L.
\end{align}
Thus, 
\begin{subequations}
\begin{align}
\fullvec{j}_{\text{iso}}^L = \dfrac{1}{3 S_0} \left[ \int d\Theta_k^M \left( \fullvec{\mathcal{S}}_{\fullvec{k}}^M \cdot \fullvec{k}^M \right) \right]
\unitvec{s}^L
\end{align}
\begin{align}
	S_0 = \int d\Theta_k \left( \left|\fullvec{D}_{\fullvec{k},\frac{1}{2}}\right|^2+ \left|\fullvec{D}_{\fullvec{k},-\frac{1}{2}}\right|^2 \right).
\end{align}
\end{subequations}
The same procedure is done for both linearly and circularly polarized light.

\end{widetext}

\bibliography{spin-current.bib}

\end{document}